\documentclass[11pt]{article}
\usepackage{psfig}

\begin{document}

\newcommand{\bra}{\langle}
\newcommand{\ket}{\rangle}
\newcommand{\beq}{\begin{equation}}
\newcommand{\eeq}{\end{equation}}
\newcommand{\bea}{\begin{eqnarray}}
\newcommand{\eea}{\end{eqnarray}}
\def\tr{{\rm tr}\,}
\def\href#1#2{#2}

\begin{titlepage}

\begin{center}
\hfill UW-PT-02/27\\
\hfill hep-th/0212041
\vspace{2cm}

{\Large\bf Lightcone Quantization of String Theory Duals of
Free Field Theories 
}

\vspace{1.5cm}
{\large
Andreas Karch }

\vspace{.7cm}

{Department of Physics,
University of Washington,
Seattle, WA 98195, USA\\
[.4cm]}
({\tt karch@phys.washington.edu}) \\

\end{center}
\vspace{1.5cm}

We quantize in light cone gauge the bosonic
sector of string theory on 
Anti-de Sitter space
in the zero curvature radius limit.
We find that the worldsheet falls apart into a theory of free partons
and map the Hilbert space of the string theory to the Hilbert space
of a free scalar in light-front description.
We outline how the string worldsheet reproduces the
field theory at weak coupling.

\vspace{3.5cm}
\begin{center}
\today
\end{center}
\end{titlepage}

\section{Introduction}
Maldacena's duality \cite{juan} was the first example of a string
theory being exactly equivalent to a gauge theory. Both sides are
governed by two parameters: one controlling the genus expansion
($1/N$ in the gauge theory and $g_s$ in the string theory) and one
governing perturbation theory for the evaluation of the
amplitude at a given genus ($\lambda = g^2_{YM} N$,
the 't Hooft coupling, and $R^4/l_s^4$, the curvature radius in string
units respectively). While the genus expansion goes parallel on both
sides, $g_s \sim 1/N$, in terms of the $\lambda$ expansion 
the gauge/gravity correspondence
is a strong-weak duality: $\lambda \sim R^4/l_s^4$, for small $\lambda$ 
the gauge theory is perturbative, for large $\lambda$ string theory
reduces to gravity. Latter fact suggests that the bulk string
theory should simplify dramatically not just in the large $\lambda$
but also in the $\lambda \rightarrow 0$ limit. After all, it is
equivalent to a free field theory.

In this note we give evidence that this is indeed the case. Going to light
cone gauge, we quantize the bosonic string, or equivalently the bosonic part
of the superstring, on AdS space in the limit where the
curvature radius in string units vanishes. We obtain the exact
spectrum of the theory. We show that the result 
agrees with what one obtains for a gauge theory in light-front
quantization: the spectrum of the string reproduces
the spectrum of free partons. The momentum in the extra dimension
encodes the fraction of lightcone momentum carried by the
individual partons.

To get a YM theory as the dual theory we
have to include the fermions on the string side.
The free, purely bosonic theory is claimed to be dual
to a matrix valued free scalar field with $U(N)$ gauge invariance.
It should also be possible to find an
example where the bosonic string has a stable highly curved AdS background
beyond zero coupling,
which then should map to an interacting scalar field theory, e.g. 
in 3 dimensions.
Such a string sized AdS seems to be a reasonable outcome of the closed
string tachyon condensation, since it at least satisfies the naive
intuition that the potential energy decreases along the condensation process.
Since we are working in light cone gauge, the conformal anomalies are quite 
subtle to detect and we hope to be able to come back to this 
issue in the future. 
As far as the superstring goes, Tseytlin
showed \cite{tseytlin} that for the full IIB string in $AdS_5 \times S^5$
the light cone worldsheet action in the $\lambda \rightarrow 0$ limit
simplifies dramatically as well. However 
the analysis won't be quite
as simple as the one we are going to present for the bosonic case.

Let us briefly comment on related work: Bardakci and Thorn
have worked out a string representation for free scalars
in \cite{bart}. It is not clear yet how they work is related to
our approach which is rooted in the AdS/CFT correspondence: their
string does not seem to know about the extra dimension. 
It would be nice to make the connection
more precise, since their approach was already extended to include
fermions and perturbation theory by construction agrees with
field theory \cite{thgauge}.
The idea that the tension of the string appears
as a perturbation in strongly curved spacetimes already appeared
in \cite{greek}.

The paper is organized as follows: in the next section we will 
quantize the bosonic string (or the bosonic sector of the superstring)
in light cone gauge on zero radius AdS and show how the theory
can be treated perturbatively in a double expansion in $g_s$ and $\lambda$.
In section 3 we give a very brief review of field theories in the
light front and show that in the free limit, the field theory
and the string theory give an identical spectrum.
In section 4 we conclude.

\section{The spectrum}
\subsection{Light Cone Gauge Hamiltonian}

Consider the bosonic part of a string moving in a warped background geometry.
The fermionic pieces of the superstring 
are not expected to alter the story significantly. Since
we are working in light cone gauge, they don't pose any conceptual problems: 
at least for type II strings we can treat them and their 
coupling to RR background
fields in
the GS formalism. Something similar should work for the non-critical 
type 0  examples as well.
The $D+1$ dimensional background metric is taken to be of the general form
\beq
ds^2 = G_{\mu \nu} 
dZ^{\mu} dZ^{\nu} = e^{2 A(Y)} 
(2 dX^+ dX^- + dX^{i} dX^{j} \delta_{i j} + dY^2)
\eeq
where $\mu$, $\nu$ run from 0 to $D$ and $i$, $j$ run over the
transverse $X$s, that is from 2 to $D-1$. We will also often
use $M$, $N$ running from $2$ to $D$ to label the
transverse $X$s and $Y$. We will be mostly interested
in AdS backgrounds with $e^{2 A(Y)} =\frac{R^2}{Y^2}$,
which describe conformal examples and should also be relevant
for the asymptotic free regime of QCD.
In the bosonic part of the string action
\beq
S= -\frac{1}{4 \pi \alpha'} \int_{M} d \tau d \sigma (-\gamma)^{1/2}
\gamma^{ab} \partial_a Z_{\mu} \partial_b Z_{\nu} G^{\mu \nu}
\eeq
one can choose the light cone gauge \cite{met1,sp}
\bea
X^+ &=& \tau \\
\gamma_{01} &=& 0 
\eea
to obtain a light cone Hamiltonian \cite{sp}
\beq
H= \frac{1}{2} \int_0^{2 \pi P_-} d \sigma \left \{ P_X P_X +
P_Y P_Y + \frac{e^{4A}}{\alpha'^2}
( X' X' + Y' Y') \right \}.
\eeq
We will mostly focus on the case of AdS where the Hamiltonian becomes
\beq
H= \frac{1}{2} \int_0^{2 \pi P_-} d \sigma \left \{ P_X P_X +
P_Y P_Y + \frac{\lambda}{ Y^4}
( X' X' + Y' Y') \right \}.
\eeq
where $\lambda=R^4/l_s^4$ measures the curvature radius in string units
and is what becomes the 't Hooft coupling of the dual gauge theory
living on the boundary of AdS as in the correspondence of \cite{juan}.

\subsection{The zero curvature radius limit}

First let us discuss the $\lambda \rightarrow 0$ limit in which
the string theory should reproduce 
free field theory in the light cone frame. The Hamiltonian
simply becomes \cite{tseytlin}
\beq
\label{hamiltonian}
H= \frac{1}{2} \int_0^{2 \pi P_-} d \sigma \left \{ P_X P_X +
P_Y P_Y \right \}.
\eeq
This is just the Hamiltonian of an infinite number of free particles.
Since all $\sigma$ derivatives dropped out of the Hamiltonian, each 
spatial mode propagates on its own, independent of the others, basically
ripping apart the string. As we will soon see, this is the stringy 
manifestation of the partons in the gauge theory!
Note that the same Hamiltonian would govern formally the $\alpha' \rightarrow
\infty$ limit in flat space string theory. However, $\alpha'$ has
dimension of length, so taking it to infinity just means we are zooming
in on the high energy states of the theory. On the contrary, in AdS we
have a dimensionless coupling constant $\lambda$. In the strict 
$\lambda \rightarrow 0$ the full string theory is captured by
the free parton Hamiltonian. Duality to weakly coupled field
theory ensures that this limit is well behaved.

There is one difference between (\ref{hamiltonian}) and $D-1$ towers
of free particles: in the original metric the range of $Y$ was between
$\infty$ and $0$, where $Y=0$ is the boundary of AdS. Even though
the divergent $1/Y^4$ factor drops out in the $\lambda \rightarrow 0$ limit,
we should still impose the finite range of $Y$. This is easiest done
by imposing an orbifold projection on Y
\beq Y \rightarrow -Y
\eeq
which just identifies negative values of Y to be positive values of $Y$.
In the end we are only going to keep states that are invariant
under the orbifold projection, that is the untwisted sector. We know
that in general modular invariance forces us to introduce in addition
a twisted sector living at the orbifold fixed plane, that is at $Y=0$.
So the twisted sector fields would look like singletons, living only
at the boundary of AdS. In the usual AdS spirit one would suspect that
they correspond to the ``decoupled center of mass motion", that is
the dynamics of the $U(1)$ part of $U(N)$. However it is not clear
that our limiting procedure is valid at $Y=0$. We will leave a detailed
analysis of the twisted sector for future work.

\subsubsection{Open String:}

For the open string a general solution to the equations
of motion satisfying Neumann boundary conditions can be written as
\beq
Z^M(\sigma, \tau) = (z^M_0 + \frac{p^M_0}{ 2 \pi P_-} \tau) \; + \;
\sum_{n > 0}
(2 z^{M}_n + \frac{p^{M}_n}{ \pi P_-} \tau ) \cos( \frac{n \sigma}{2 P_-} ) .
\eeq
The canonical momentum is
\beq
P_M(\sigma,\tau) = \dot{Z}^M = \frac{p^M_0}{2 \pi P_-}
\; + \;  
\sum_{n > 0}
\frac{p^{M}_n}{ \pi P_-}  \cos( \frac{n \sigma}{2 P_-} ). 
\eeq
Imposing canonical commutation relations on $Z^M(\sigma,\tau)$ 
and $P_M(\sigma',\tau)$ shows that all the $z^M_n$, $p^M_n$ are canonical
coordinates and momenta themselves
\beq
[z_n^M, p_m^N] = i \delta_{nm} \delta^{MN}.
\eeq
Plugging back into the Hamiltonian (\ref{hamiltonian}), we indeed
recover just a collection of free particles
\beq
2 \pi H=\frac{1}{P_-} (p^M_0)^2 \; + 
\; \frac{1}{2 P_-} \sum_{n > 0,M} (p_n^M)^2.
\eeq
So the states are just given by a tensor product of free particle 
momentum eigenstates, that is they
are of the form
\beq
\left | \left \{ (p^y)_0, (p^{\perp})_0 \right \}; \, \ldots \, ; 
\left \{ (p^y)_n, (p^{\perp})_n \right \} ; \, \ldots \right  > 
\eeq
where $(p^y)_n$ and
$(p^{\perp})_n$ are the $y$ and transverse momentum of the particles
respectively.
Under the orbifold
action all
\beq
(p^y)_n \rightarrow - (p^y)_n,
\eeq 
so to impose the orbifold 
constraint we should only allow invariant states of the form
\beq 
\left | \ldots; \, \left \{ (p^y)_n, (p^{\perp})_n\right \};
 \, \ldots \right > \; \;  + \; \;  
\left | \ldots; \, \left \{ -(p^y)_n, (p^{\perp})_n \right \} ;
 \, \ldots \right >
\eeq
where we should now restrict to
\beq
 p^y_n \geq 0 
\eeq
in order to avoid redundancy. 

\subsubsection{Closed String:}

For the closed string a general periodic solution to the equations
of motion can be written as
\beq
Z^M = 
 \sum_{n=-\infty }^{\infty}
\left [ 
(z^{M}_n + \frac{p^{M}_n}{2 \pi P_-} \tau ) e^{ \frac{ i n \sigma}{P_-}}
\right ]
\eeq
where reality demands $p_{-n}^{\dagger} = p_n$
and again we obtain a collection of free particles
\beq
  2 \pi H= \frac{1}{P_-} \sum_{n,M} p_{-n}^M \, p_n^M =
 \frac{1}{ P_-} (p_0^M)^2  \; + \;
 \frac{2}{ P_-} \sum_{n >0  ,M} |p_n^M|^2.
\eeq
For the closed string we have to impose in addition the zero
momentum constraint in order to insure gauge invariance under
$\sigma$ translations
\beq
\label{gauss}
0 = P = \int_0^{2 \pi P_-} d \sigma \sum_M P_M (Z^M)' =
\sum_{n,M} \frac{i n}{P_-} \left ( 
x^M_n p^M_{-n} \right )
\eeq
The orbifold action, as in the open case, just restricts all the $p_y$
to be non-negative. 

\subsection{Interactions}

So far we have only quantized the theory in the $g_s=0$, $\lambda=0$
limit. But it is obvious that perturbation theory in those two 
parameters is conceptually straight forward. The $\lambda$
perturbation theory just amounts to good old fashioned
quantum mechanical perturbation theory with the interaction
Hamiltonian
\beq
H_1 = \lambda \int_0^{2 \pi P_-} d \sigma \frac{1}{Y^4} \left ( (X')^2 + (Y')^2 
\right ).
\eeq
In order to reproduce the $g_s$ expansion we have to allow splitting and 
joining of strings. Given the simplicity of our Hamiltonian the
corresponding string field theory vertex should be not too hard to construct. 
Note that the $\lambda$ perturbation theory becomes singular
whenever $Y$ tends to zero. Some regularization procedure is necessary.
One way would be to cut out regions from the worldsheet where
$Y<\epsilon$. Regularization is to be expected once 
we are describing the string
theory dual to an interacting theory like 
YM: infinities arise and lead to dimensional transmutation
of the coupling constant. In the case of ${\cal N}=4$ SYM all singularities
in the $\epsilon \rightarrow 0$ limit have to cancel between the bosons and
fermions. We want to emphasize again that the dual field theory makes
it clear, that the $\lambda \rightarrow 0$ limit is smooth.

\section{The dual field theory interpretation}

\subsection{Light Cone Quantization in Field Theory}
The holographic dual is a free $D$-dimensional field
theory of $N \times N$ matrices with
$U(N)$ gauge invariance, that is we restrict physical observables
to be $U(N)$ invariant. For the open string case there are
in addition $M$ fundamental matter multiplets, where $M$ denotes
the dimension of the Chan-Paton space (the number of spacetime filling
D-branes). For the superstring case, this has to be augmented by
free adjoint gauge fields, 
fermions and additional scalars, say to get ${\cal N}=4$ SYM, but
the bosonic sector we quantized above is already supposed to be dual to a
the free scalar matrix valued field with gauge invariance.

Quantizing the string theory in the light-cone gauge forces us
onto the light cone on the field theory side as well. So we briefly
need to review the structure of YM theory in light cone quantization
\cite{kogsus}, see \cite{brodsky} for a recent review.
As is known
to most string theorists from the study of (M)atrix theory \cite{bfss},
light cone quantization leads to a non-relativistic structure
with $P_-$ playing the role of mass in the Hamiltonian.
In particular the vacuum is trivial. It's whole complication
gets hidden in the dynamics of the zero modes.

Let's look at the quantization of a massless 
scalar field in the free limit. 
The light cone time translation generator
becomes 
\beq
P_+ = \frac{1}{2} \int dx_+ dx^{D-2}_{\perp} \; \left (
\Phi_i (i \partial_{\perp})^2 \Phi^i \right ).
\eeq
where $i$ is an adjoint $U(N)$ color index.
One obtains a collection of free particles.
One uses as a basis of states the free Fock space 
forming the eigenstates of the free Hamiltonian
\beq
\left | n: k_a^{i,+}, k_a^{i,\perp} \right >
\eeq
where $n$ denotes the
number of particles in the Fock space and the $k_a$ are the
momenta of the $a$-th particle. 
Usually one is only interested in gauge
invariant states, but we kept the color label in order to distinguish
partons with different contractions. At large
$N$ the single trace states get singled out.
It is a key property that all $k_a^+$ are positive
\beq
k_a^+ \geq 0
\eeq
since this implies the triviality of the vacuum, modulo
the complications with the zero mode.
The total $P_+$ in a Fock space state is given by a sum over
all the partons, each of which is on shell
\beq
P_+ = \sum_a \frac{(k^{\perp}_a)^2}{k^+_a}.
\eeq

We will
treat the zero mode in a somewhat unusual fashion that resembles the
treatment of soft quanta in standard perturbation theory.
The standard light-front method would be to
discretize the momenta and then to solve for the zero mode.
In this case the ground state of the
Fock space is the unique vacuum of the theory. The particles
forming the Fock space are usually referred to
as {\it partons} and it is precisely in this
language that one can see the partonic behavior
of QCD at high energies: any hadron can
be written as a wavefunction in the Fock-space
basis, and at high energies the parton constituents
become the scattering centers.
The light cone Hamiltonian
can be numerically diagonalized in the interacting theory by truncating
the Fock space. The resulting spectrum gives the masses of the bound states.
This program has been carried out with some success in QCD \cite{brodsky}.

Instead, in order to incorporate all zero momentum quanta, 
we want to focus on the special subset of parton states of the form
\beq
\tr \left | N: (k_a^+,k_a^{\perp}) \right >
\eeq  
The claim is that these span the complete physical
meaningful Hilbert space of
single trace states. If $N-k$ of the momenta go to zero (the partons
are on-shell, but since we are dealing with massless excitations the
momenta can vanish), the state looks like a $k$ parton state together
with several ``soft photons'' (wee partons). Replacing the $k$ parton state
with the full $N$ parton state is the analog of always calculating
inclusive cross-section with an arbitrary number of soft photons in order
to get well defined cross-sections. States with more than $N$ partons
can be rewritten as a multi-trace state due to $U(N)$ identities.

\subsection{Map between String Theory and Field Theory}
It follows that the spectrum we found on the string
theory side precisely reproduces the $U(N)$ invariant single-trace spectrum 
of the
free, massless scalar field 
theory, if we encode the
momentum in the ``5th dimension'' (the
$Y$ direction) of the $n$-th particle
in the string Fock space in the $k^+$ of the $n$-th
particle in the light-cone Fock space, as we will do in detail below.
The closed string maps to $N$ gluon states,
\beq
\tr|ggggg\ldots ggggg>.
\eeq
Including fundamental matter, color
singlets involving two fundamentals and $N$ adjoints,
\beq
|qgggg\ldots gg\bar{q}>,
\eeq
map to the open string spectrum.
Multitrace states are, as usual, interpreted
as multi-string states. Note that this implies that in the
string theory at finite
coupling the number of independent free particles has to
truncate at $N=\frac{1}{g_s}$, that is it is a
non-perturbative effect. This is the usual stringy
exclusion principle \cite{stringy}.

Actually, at large $N$ one would expect the partons to be ordered
along $\sigma$, whereas the free string modes we found are 
momentum modes in $\sigma$. For the transverse
momenta the right identification seems to be
to identify the momenta on the string side as waves on the single
trace $N$ parton
states, similar to the proposal of \cite{BMN}. That is
\beq
\left |0,0,\ldots,0,(p^{\perp})_n,0,\ldots,0 \right > = 
\frac{1}{\sqrt{N}} \sum_{a=1}^N
e^{\frac{2 \pi i a n}{N}} \tr  \left | 0,0 \ldots, 0,k^{\perp}_a,0, \ldots,
0 \right > .
\eeq
This actually vanishes by cyclicity of the trace, the string
state does not satisfy the zero momentum constraint (\ref{gauss}).
But in the same spirit
we can identify
\beq
\left |0,0,\ldots,0,p^{\perp}_n,0,\ldots,0, p^{\perp}_m,0, \ldots,0 \right > = 
\frac{1}{\sqrt{N}} \sum_{a,b=1}^N
e^{\frac{2 \pi i (a n+ bm)}{N}}
\tr  \left | 0,0 \ldots, 0,k^{\perp}_a,0, \ldots,
0,k^{\perp}_b,0, \ldots,0 \right > .
\eeq
Cyclicity of the trace 
means that the right hand side is invariant under translations in $\sigma=
a/N$. This translation invariance was the origin of (\ref{gauss}) and
hence we precisely reproduce the closed string spectrum,
as long as $p_y$ is given by such a discrete Fourier transform
as well, as we will find it to be the case below. 
The partons of the field theory
become string bits \cite{thorn1,thorn2} or a gluon chain \cite{greensite}. 
The open spectrum
works in an analogous fashion.

The map between
the momentum in the 5th dimension and the $k^+$ carried by
the individual partons is more subtle. Note that by group theory we only know 
that the total momenta $P_+$, $P_{\perp}$ and $P_-$ 
have to map into each other. The counting of parameters
works nicely: string theory has for each mode
 $p^{\perp}_n$ and $p^y_n$, $P_-$ has no modes by 
gauge choice and the $p^-_n$ are solved for by the constraints. In field
theory 
we have $k^{\perp}_a$ and $k^+_a$, while $k^-_a$ gets solved for
as $(k^{\perp}_a)^2/k^+_a$. The total transverse momentum
\beq
P_{\perp} =  p^0_{\perp} = \sum_a k^a_{\perp}\eeq
matches trivially. $P_-$ is just a parameter on the string side,
it matches on the field theory side to 
\beq P^+ = \sum_a k^+_a. \eeq
The map between $p^y_n$ and $k^+_a$ has to yield the
matching of $P_+$.
In detail, writing $X^-(\sigma,\tau)$ in the
same form as the transverse coordinates,
solving the constraints yields for the modes
\beq (p^-)_n = \frac{1}{2 \pi P_-} \sum_{m,M} p^M_m p^M_{n-m}. \eeq
These should be related to the $(k^-)_a = \frac{(k^{\perp}_a)^2}{k^+_a}$
by the Fourier transform
\beq
p^-_n = \frac{1}{\sqrt{N}} \sum_a e^{\frac{2 \pi i a n}{N}} k^-_a=
\frac{1}{\sqrt{N}} \sum_a e^{\frac{2 \pi i a n}{N}} 
\frac{(k^{\perp}_a)^2}{k^+_a}.
\eeq
Requiring those two expressions for $p^-_n$ to agree yields the
following map
for $p^y_n$:
\beq p^y_n = \frac{1}{\sqrt{N}} \sum_a e^{\frac{2 \pi i a n}{N}} 
\alpha_a \eeq
where 
\beq
\alpha_a = |k^{\perp}_a| \left ( \frac{1}{x_a} -1 \right  )^{\frac{1}{2}}
\eeq
and $x_a$ denotes the $P^+$ momentum fraction carried by
the $a$-th parton, 
\beq
x_a =\frac{ k_a^+}{P^+}.
\eeq 
$x$ takes values between 0 and 1, so the $\alpha_a$ take values
between $+\infty$ and $0$ just as we expect from the dual to $p_y$.
A consistent picture emerges.

\section{Conclusions}

We quantized the bosonic string on AdS backgrounds 
in light-cone gauge in the zero curvature
radius limit and obtained the exact spectrum. The string
falls apart into free partons, which map precisely to what
one would expect from a free field theory with $U(N)$ gauge
invariance in the light front formalism. 
In order to promote the field theory to an interacting theory, 
one possibility
is to add in the fermions and study the full AdS$_5$ $\times$ $S^5$
background of IIB around the zero radius limit in the GS formalism in the
same spirit. The bosonic excitations will again show the partonic
behavior of the dual theory. Another possibility that has to be
explored is that maybe substringy AdS spaces are good backgrounds
for the bosonic string in lower dimensions.

\noindent{\large\bf Acknowledgment} I would like to thank 
Joe Polchinski for drawing my attention to the light cone quantization
approach and for criticizing several misconceptions in an earlier
draft. Many thanks also to Josh Erlich, Ami Katz,
Pavel Kovtun, Gerry Miller, Steve Sharpe,
Larry Yaffe and in particular Steve Ellis
for helpful discussions. This work was partially supported by the DOE
under contract DE-FGO3-96-ER40956.

\bibliography{warpstring}
\bibliographystyle{ssg}
\end{document}